# Effects of biaxial strain on the improper multiferroicity in h-LuFeO$_3$ films

Kishan Sinha[1], Yubo Zhang[2], Xuanyuan Jiang[1], Xiao Wang[3], Xiaozhe Zhang[1,4], Philip J. Ryan[5], Jong-Woo Kim[5], John Bowlan[6], Dmitry A Yarotski[6], Yuelin Li[5], Anthony D. DiChiara[5], Xuemei Cheng[3], Xifan Wu[2]*†, Xiaoshan Xu[1,7]*‡

[1]Department of Physics and Astronomy, University of Nebraska, Lincoln, Nebraska 68588, USA,

[2]Department of Physics, Temple University, Philadelphia, Pennsylvania 19122, USA

[3]Department of Physics, Bryn Mawr College, Bryn Mawr, Pennsylvania 19010, USA

[4]Department of Physics, Xi'an Jiaotong University, Xi'an 710049, China

[5]Advanced Photon Source, Argonne National Laboratory, Argonne, Illinois 60439, USA

[6]Center for Integrated Nanotechnologies, Los Alamos National Laboratory, Los Alamos, New Mexico 87545, USA

[7]Nebraska Center for Materials and Nanoscience, University of Nebraska, Lincoln, Nebraska 68588, USA

*To whom all correspondence should be addressed.

† xifanwu@temple.edu; ‡ xiaoshan.xu@unl.edu

Abstract

Elastic strain is potentially an important approach in tuning the properties of the improperly multiferroic hexagonal ferrites, the details of which have however been elusive due to the experimental difficulties. Employing the method of restrained thermal expansion, we have studied the effect of isothermal biaxial strain in the basal plane of h-LuFeO$_3$ (001) films. The results indicate that a compressive biaxial strain significantly enhances the ferrodistortion, and the effect is larger at higher temperatures. The compressive biaxial strain and the enhanced ferrodistortion together, cause an increase in the electric polarization and a reduction in the canting of the weak ferromagnetic moments in h-LuFeO$_3$, according to our first principle calculations. These findings are important for understanding the strain effect as well as the coupling between the lattice and the improper multiferroicity in h-LuFeO$_3$. The experimental elucidation of the strain effect in h-LuFeO$_3$ films also suggests that the restrained thermal expansion can be a viable method to unravel the strain effect in many other epitaxial thin film materials.

In an improperly ferroelectric material, the ferroelectric order can be induced by a ferrodistortion [1], while in a weakly ferromagnetic material, the ferromagnetic order is typically caused by a structural distortion [2]. Therefore, in materials that are both improperly ferroelectric and weakly ferromagnetic, the structural distortions can play a critical role both in originating the ferroic orders and in coupling them. As a prototypical improper ferroelectric material, hexagonal $LuFeO_3$ (h-$LuFeO_3$) exhibits ferroelectricity below 1050 K and weak ferromagnetism below 130 K. [3–5] The ferroelectricity is induced by a ferrodistortion ($K_3$ structural distortion, see Fig. 1(a)). [5–8] The ferromagnetic order, which is parasitic to the antiferromagnetic order, is induced by the $K_3$ ferrodistortion both in terms of the creation of the local magnetic moments by the Dzyaloshinskii-Moriya (DM) interactions [9,10] and in terms of the non-zero inter-layer exchange interactions due to the reduction of symmetry. [5,6,11] First principle calculations predict a possible reversal of magnetization by an electric field along the *c* axis [6], and anomalously large magnetoelectric effects in the *a-b* plane in h-$LuFeO_3$ [12], both of which are mediated by the lattice.

Experimentally, detailed roles of the ferrodistortion can be studied by varying the distortion, often by applying an elastic strain. Elastic strain is a promising tool for studying and tuning material properties, such as ferroelectricity, magnetism, catalysis, and transport properties [13–17], in addition to the methods such as chemical strain or altering the material structures, owing to the universal coupling between the crystal structure and electronic structures in materials [18,19]. This is unfortunately difficult for h-$LuFeO_3$, which is unstable in bulk but can be stabilized in epitaxial thin films: The lack of structurally compatible substrates makes the growth of defect-free films impossible and makes the epitaxial strain difficult to control [4,5,20] and there are no bulk counterparts to compare with since the stand-alone hexagonal phase of $LuFeO_3$ is unstable. As a result, investigations on the strain effect in h-$LuFeO_3$ have been rare.

In this work, we employed a method of restrained thermal expansion to study the strain effect in h-LuFeO$_3$. In general, thermal strain may be generated in a material in all crystalline dimensions in an isobaric thermal expansion. However, in an epitaxial thin film, if the film is selectively heated, the out-of-plane dimension is free to expand while the in-plane dimensions of the film is restrained by the substrate (restrained thermal expansion) [Fig. 1(b)]. By comparing the material properties in the isobaric and restrained thermal expansions, the effect of isothermal compressive strain at a higher temperature can be obtained [Fig. 1(c)] (see also S1 in the supplementary materials) [21]. Using the method of restrained thermal expansion, we studied the strain effect on the K$_3$ ferrodistortion in h-LuFeO$_3$. Experimentally, we observed that the biaxial strain in the basal plane of h-LuFeO$_3$ does significantly couple to the K$_3$ ferrodistortive, which agrees with the results of our first principle calculation. We also found from our first principle calculations that the compressive strain combined with the enhanced ferrodistortion, increases the spontaneous electric polarizations but reduces the weak ferromagnetic moments.

Hexagonal LuFeO$_3$ (001) thin films (25 nm) were grown using pulsed laser deposition on Al$_2$O$_3$ (001) substrates. [4,11,22] In the restrained thermal expansion, the thermal strains were measured in a laser-pump x-ray-probe style [23–26], as shown in Fig. 1(d). Short durations of restrained thermal expansion in the h-LuFeO$_3$ thin film, were generated using laser pulses (30 ps, 1 kHz) with the photon energy (3.2 eV) between the band gap of the Al$_2$O$_3$ substrate (8.8 eV) [27,28] and that of the h-LuFeO$_3$ film (2.0 eV) [22,29]. Time resolved diffractions were carried out on h-LuFeO$_3$ (106) peaks to measure the lattice constants and the structural distortions, using X-ray pulses (80 ps, 12 keV), at different time delay ($\Delta t$) with respect to the laser pulses, with a two-dimensional detector. The temperature dependent X-ray diffractions were carried out to measure the isobaric thermal expansion between 20 and 485 K at the beamline 6-ID-B, and the time-resolved X-ray diffractions were carried out at the beamline 14-ID-B of the Advanced Photon Source at the Argonne National Laboratory. Our first-principles calculations are carried based on density functional theory as implemented in the Vienna ab initio simulation package (VASP) [30]. We adopt the Perdew-Burke-Ernzerhof functional revised for solid (PBEsol) [31] where the spin-polarized generalized gradient approximation (GGA) is made in treating the exchange correlation of the electrons. A cutoff energy of 500 eV is used in the plane-wave basis with a 4×4×2 k-point mesh centered at Gamma point. For transition metals, we choose $U$=4.5 eV and $J_H$=0.95 eV. The criterion of residual Hellman-Feynman forces for structural relaxation is 0.001 eV/Å.

First, we demonstrate that the thermal strain only occurs in the out of plane direction in the restrained thermal expansion. Figure 2(a) shows the scans of the (106) peak before ($\Delta t < 0$) and after the laser pulse ($\Delta t = 0.2$ ns) of a 1.0 mJ/mm$^2$ fluence. The lattice of the h-LuFeO$_3$ film at room temperature (without the laser illumination) is used as the reference coordinate system. A clear shift of the diffraction profiles in the reciprocal index $L$ is observed in Fig. 2(a), indicating a thermal strain along the $c$ axis. In contrast, there is no observable strain along the $a$ axis [Fig. 2(b)], suggesting that the in-plane axis is restrained by the Al$_2$O$_3$ substrate that has no thermal strain because its band gap is too high to absorb the laser photon.

The change of the K$_3$ ferrodistortion can be estimated from the intensity change of the (10$L$) peaks, because the K$_3$ ferrodistortion is directly related to the diffraction intensity of (10$L$) peaks as $I_{(10L)} \propto Q_{K3}^2$ [4] if the contribution of the oxygen is ignored, where $Q_{K3}$ is the amplitude of the K$_3$ ferrodistortion. As shown in Fig. 2(b), the (106) peak intensity decreases as the laser pulse heats the film, which is expected, because the (10$L$) peak vanishes at the ferroelectric → paraelectric transition at high temperature. [4]

Next, we show that the observed time evolution of the thermal strain and the diffraction peak intensity can be explained in terms of thermal conduction. As shown in Fig. 2(b), both the change of peak intensity and the thermal strain decay over time with a similar trend. In the case of thermal conduction, the temperature in the film follows the diffusion equation $\rho c_M \left(\frac{\partial T}{\partial t}\right)_z = \sigma \left(\frac{\partial^2 T}{\partial z^2}\right)_t$, where $\rho$, $c_M$, $T$, $t$, $z$, and $\sigma$ are the mass density, the mass specific heat, temperature, time, direction of the thermal conduction, and thermal conductivity respectively. [23,32] At the film/substrate interface, the diffusion equation becomes $\rho c_M \left(\frac{\partial T}{\partial t}\right)_z = g\Delta T$, where $g$ and $\Delta T$ are the interfacial thermal conductivity and the temperature difference at the interface respectively. The thermal strain $\frac{\Delta c}{c}$ is expected to decay in a similar trend. The only unknown parameter here is the interficial thermal conductivity $g$. As shown in Fig. 2(b), we fit the time dependence of the thermal strain $\frac{\Delta c}{c}$ with the diffusion equations using $g$ as the fitting parameter. The result shows $g = 3.8 \times 10^8$ W/(K m$^2$), which falls into the proper range of the thermal conducitivty of the epitaxial interfaces. [33,34] The fact that the decay of thermal strain can be explained in the light of thermal conduction, suggests that temperature is a well-defined state function during the decay process (see also S2-S5 in the supplementary materials [21]).

Because temperature is well defined in the decay process, one may calculate the isothermal strain effect by comparing the properties in the isobaric and restrained thermal expansions at the same temperature [Fig. 1(c)]. The thermal strain in the isobaric thermal expansion, is displayed in Fig. 2(c). As the temperature increases, both $a$ and $c$ axes expand. The linear thermal expansion coefficients of the h-LuFeO$_3$ film around room temperature in both $a$ and $c$ directions are found to be $(8.0 \pm 0.1) \times 10^{-6}$ (see also S6 in the supplementary materials [21]). The intensity of the (106) peak also decreases as the temperature increases [Fig. 2(d)] in the isobaric thermal expansion, as it does in the restrained thermal expansion.

Using the thermal and optical properties of h-LuFeO$_3$, [22,29,35] we estimated that the temperature change of the h-LuFeO$_3$ film after absorbing a 1.0 mJ/mm$^2$ photon pulse to be ~ 460 K (±10%) (see also S7 in the supplementary materials [21]). The relation between the thermal strain $\frac{\Delta c}{c}$ and the fluence in Fig. 2(b) inset, can be converted to the relation between $\frac{\Delta c}{c}$ and temperature, which is used to estimate the temperature in the restrained thermal expansion. In Fig. 3(a), $Q_{K3}$ in the restrained thermal expansion, is calculated according to $I_{(10L)} \propto Q_{K3}^2$ and the Debye-Waller factors (see also S8 in the supplementary materials [21,36]), and plotted against the temperature. Also plotted is the temperature dependence of $Q_{K3}$ in the isobaric thermal expansion calculated according to the data in Fig. 2(d). Obviously, $Q_{K3}$ is enhanced in the restrained thermal expansion.

In order to find the effect of isothermal strain on another physical property (e.g. amplitude of K$_3$ ferrodistortion $Q_{K3}$), one needs to compare the temperature dependences of the physical property in the isobaric and restrained thermal expansions, as depicted in Fig. 1(c). In the restrained thermal expansion, the change of a general physical property (state function) $f$, relative to an initial state $(a_0, T_0, f_0)$, can be written as $f_a - f_0 \approx \left(\frac{\partial f}{\partial T}\right)_{a,\sigma_c} \Delta T$, where $\sigma_c$ is the stress along the $c$ axis. In the isobaric thermal expansion, the change of the physical properties corresponds to $f_P - f_0 \approx \left(\frac{\partial f}{\partial T}\right)_{\sigma_a,\sigma_c} \Delta T$. The two processes can be related using Legendre transformation and chain rules of partial differential [32]: $\left(\frac{\partial f}{\partial T}\right)_{\sigma_a,\sigma_c} = \left(\frac{\partial f}{\partial T}\right)_{a,\sigma_c} + \left(\frac{\partial f}{\partial a}\right)_{T,\sigma_c}\left(\frac{\partial a}{\partial T}\right)_{\sigma_a,\sigma_c}$, where $\left(\frac{\partial f}{\partial a}\right)_{T,\sigma_c}$ describes the isothermal strain effect, which can be found as $\left(\frac{\partial f}{\partial a}\right)_{\sigma_c,T} \approx -\frac{f_a-f_P}{\Delta T}\frac{1}{\left(\frac{\partial a}{\partial T}\right)_{\sigma_a,\sigma_c}}$. As shown in Fig. 1(c), if the state after the isobaric thermal expansion is used as the reference, the strain can be defined as $\frac{\Delta a}{a} \equiv -\frac{1}{a}\Delta T\left(\frac{\partial a}{\partial T}\right)_{\sigma_a,\sigma_c}$ (compressive); the change of $f$ caused by the strain is $f_a - f_P$. Figure 3(b) shows the relation between $Q_{K_3,a} - Q_{K_3,P}$ and the in-plane biaxial strain $\frac{\Delta a}{a}$. The data points indicate the strain effect at certain temperature (top axis), measured at a certain magnitude of the strain (bottom axis). Obviously, the isothermal compressive biaxial strain enhances the $K_3$ lattice distortion. In addition, the effect of strain on $Q_{K_3}$ appears to be larger at higher temperatures.

To better understand the effect of strain on the $K_3$ ferrodistortion experimentally measured by the method of restrained thermal expansion, we carried out first-principles calculations based on density functional theory to elucidate the structural distortions at the atomic level. The structures are fully relaxed, for h-LuFeO$_3$ of the space group symmetry P$_{63}$cm under the epitaxial strains ranging from -2% to 2%. Based on the relaxed structures, the mode decompositions were performed using the group theory. The resulting $Q_{K3}$ are presented in Fig. 4(a) as a function of biaxial strain. Indeed, our theoretical calculations show that $Q_{K3}$ is enhanced (reduced) by the applied compressive (tensile) epitaxial strains, which is consistent with the experimental observation. Under the compressive strain, all the atoms are forces to be more compactly packed within the unit cell. As a result, the intralayer Fe-O bond lengths are slightly reduced and those between Fe and apical oxygen atoms are slightly increased. In addition, the Fe-O bonds within the trimer structure also respond by a bucklering behavior compatible with the $K_3$ ferrodistortion. As schematically shown in Fig. 1(a), the oxygen atom at the center of the trimer which is shared by three bi-pyramids is moving up, while the other oxygen atoms in the bases of the three bi-pyramids are all moving downwards. As a result, $Q_{K3}$ is increased. At the same time, the intralayer distances between two neighboring Fe or Lu atoms are reduced due to the compressive strain.

It is well known that the electric polarization is strongly coupled to the epitaxial strain in properly ferroelectric materials such as BaTiO$_3$. Yet, the coupling between the epitaxial strain and functional properties in multiferroic materials has been much less addressed. We next focus on the tunabilities of functional properties in h-LuFeO$_3$, including electric polarization and weak ferromagnetism, under the epitaxial strains. As an improperly multiferroic material, similar to YMnO$_3$ [8], the ferroelectric distortion in h-LuFeO$_3$ is driven improperly by the K$_3$ ferrodistortion that can be described by the rotation of FeO$_5$ trigonal bi-pyramids and the buckling of Lu layers [Fig. 1(a)], which is a highly unstable structural instability in its centrosymmetric P$_{63}$/mmc phase. [5–8] Therefore, it is expected that the polarization should increase as $Q_{K3}$ increases. [7] Indeed, as shown in Fig. 4(b), the polarization is enhanced linearly under the compressive biaxial strain. The change of polarization has a similar rate to that of $Q_{K3}$. The tunability of polarization by epitaxial strain is much less than that of the conventional ferroelectric materials such BaTiO$_3$. It indicates that the piezoelectricity is relatively small which is consistent with recent experiment in improper hexagonal YMnO$_3$ [37]. The weak ferromagnetism originates from both the DM interaction and single ion anisotropy. The magnitude of the DM interaction dependents on the DM vector $\mathbf{D} \sim |\mathbf{r}_{Fe-Fe} \times \boldsymbol{\delta}_z|$ [6,9,10], where $\mathbf{r}_{Fe-Fe}$ is the displacement vector between the two iron atoms and $\boldsymbol{\delta}_z$ is the displacement vector along [001] direction for the oxygen atom shared by three bi-pyramids in the trimer shown in Fig.1(a) respectively. Since $\boldsymbol{\delta}_z$ is closely associated with the trimerization measured by the $Q_{K3}$, the weak ferromagnetism was found to be intrinsically related to the K$_3$ ferrodistortion. Therefore, an enhanced ferromagnetic moment is expected at a larger $Q_{K3}$ under compressive strain. However, the compressive strain also brings the two Fe atom closer which reduces the displacement vector $\mathbf{r}_{Fe-Fe}$ more rapidly [see Fig. 1(a)]; this actually reduces amplitude of cross product of the DM vector. As a result, the canting ferromagnetic moment is rather decreased under compressive biaxial strain [see Fig. 4(b)].

In conclusion, by the combined experimental and theoretical studies, we have found a significant coupling between the biaxial strain in the basal plane and the $K_3$ ferrodistortion in h-LuFeO$_3$, which in turn couples to the electric and magnetic polarizations in this improperly ferroelectric and weakly ferromagnetic material. In particular, the compressive strain enhances the $K_3$ ferrodistortion and the ferroelectric polarization, but reduces the canting of weak ferromagnetic moments. The elucidation of the strain effect in h-LuFeO$_3$ is an important advancement of our understanding on the coupling between the lattice and the improper multiferroicity. The experimental characterization of strain effect in h-LuFeO$_3$, can potentially be extended to measure the electronic and magnetic properties, when additional probes (e.g. optical or soft X-ray) are included. This could be especially important for studying the epitaxial thin films for which the strain effects haven't been fully investigated due to the imperfection in epitaxy or the lack of bulk counterparts.

## Acknowledgement

The experimental effort in this work was mainly supported by the National Science Foundation (NSF), DMR under Award DMR-1454618. X.M.C. acknowledges partial support from National Science Foundation Grant No. DMR-1053854. The theoretical effort was supported by the Air Force Office of Scientific Research under FA9550-13-1-0124. This research used resources of the Advanced Photon Source, a U.S. Department of Energy (DOE) Office of Science User Facility operated for the DOE Office of Science by Argonne National Laboratory under Contract No. DE-AC02-06CH11357. Use of BioCARS was also supported by the National Institute of General Medical Sciences of the National Institutes of Health under grant number R24GM111072. The content is solely the responsibility of the authors and does not necessarily represent the official views of the National Institutes of Health. Time-resolved set-up at Sector 14 was funded in part through a collaboration with Philip Anfinrud (NIH/NIDDK).

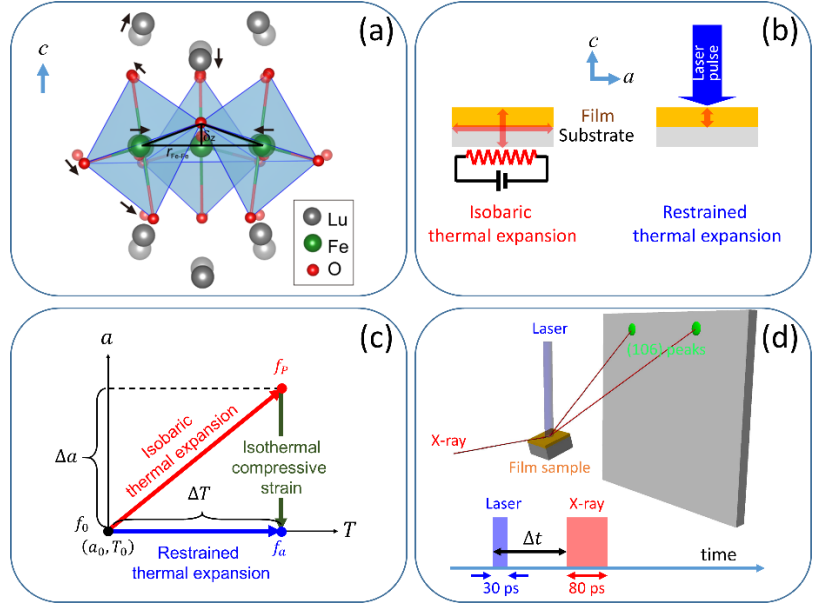

**Figure 1**. (color online) (a) Trimer model of the $K_3$ ferrodistortion in h-LuFeO$_3$, which corresponds to the rotation of FeO$_5$ trigonal bipyramids, the buckling of the Lu layers, and the trimerization. The arrows indicate the atomic displacements when $K_3$ is enhanced under the compressive biaxial strain. (b) Schematics of the strain generated by the restrained thermal expansion in comparison with that by the isobaric thermal expansion. The arrows indicate the directions of thermal expansion. (c) Illustration of the restrained thermal expansion and isobaric thermal expansion in the ($a$, $T$) space, where $a$ is the in-plane lattice constants, $T$ is temperature, $f$ is a general physical property. (d) Illustration of the experimental setup for the pump (laser) and probe (x-ray) measurements.

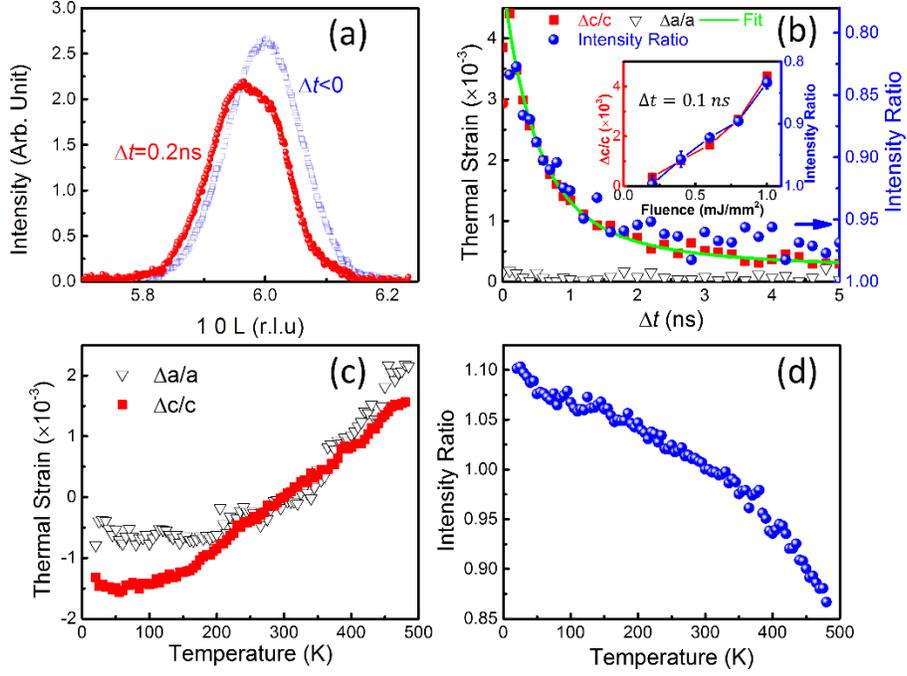

**Figure 2**. (color online) Thermal strain and the intensity change of the (106) peak in the restrained and isobaric thermal expansions. (a) The diffraction profile of the (106) peak before and after the laser (1.0 mJ/mm$^2$) pulse. (b) The decay of the thermal strain $\frac{\Delta c}{c}$ and the (106) peak intensity in the restrained thermal expansion, as well as the fit using the thermal conduction model. Inset: the thermal strain $\frac{\Delta c}{c}$ and the (106) peak intensity at $t_{delay} = 0.1$ ns as a function of laser fluence. (c) and (d) are the thermal strain and (106) peak intensity respectively, as a function temperature in the isobaric thermal expansion. The (106) peak intensities are normalized using the values at room temperature.

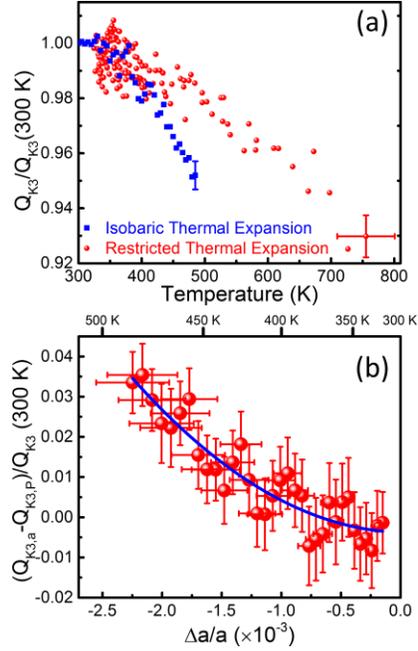

**Figure 3**. (color online) (a) Temperature dependence of the amplitude of the K$_3$ distortion ($Q_{K3}$) in both isobaric and restrained thermal expansion as a function of temperature. Representative error bars are displayed. (b) The effect of isothermal biaxial compressive strain on $Q_{K3}$. Each data point represents a change of $Q_{K3}$ caused by a strain $\frac{\Delta a}{a}$ (bottom axis) at the corresponding temperature (top axis). The line is a guide to the eyes.

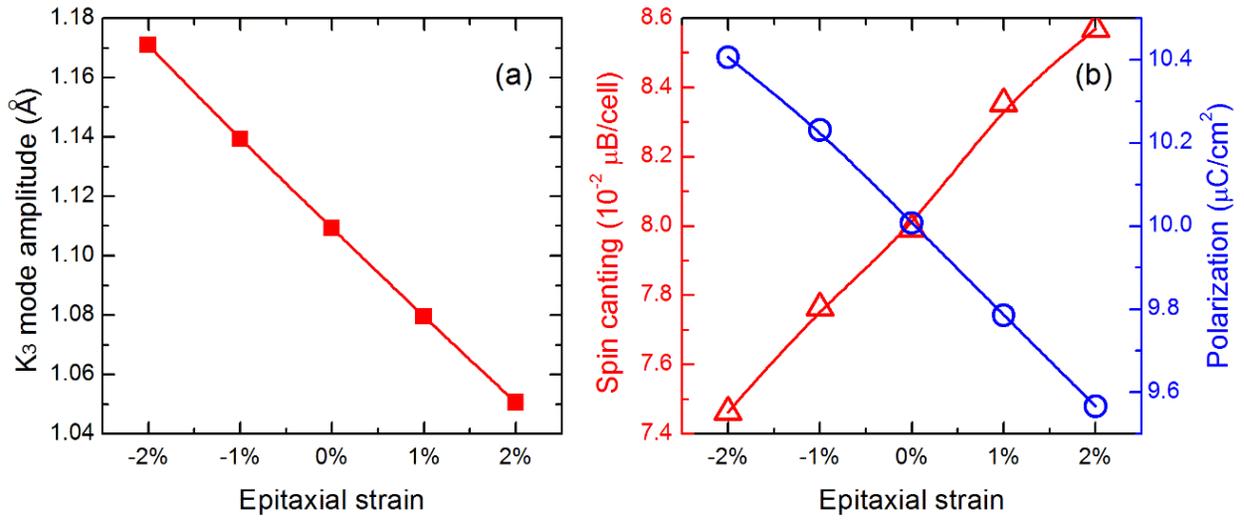

**Figure 4**. (color online) Effect of the biaxial strain $\frac{\Delta a}{a}$ on the $K_3$ distortion (a) and on the electric polarization and the weak ferromagnetic moment, calculated using the density functional theory.

# Effects of biaxial strain on the improper multiferroicity in h-LuFeO$_3$ films: Supplementary materials

## S1. Two-process method for measuring the effect of epitaxial strain

Upon isobaric thermal expansion, thermal strain may be generated in a material in all crystalline dimensions. In an epitaxial thin film, if the film is selectively heated, the in-plane dimensions ($a$) of the film is restrained by the substrate, while the out-of-plane dimension ($c$) is free to expand (restrained thermal expansion). By comparing the material properties in the isobaric and restrained thermal expansions, the effect of isothermal compressive strain at a higher temperature can be obtained (Fig. S1). In this method, the requirement on epitaxy is less stringent and the effects of extrinsic factors introduced by comparing samples grown on different substrates may be minimized. In order to show how to measure the effect of epitaxial strain, here we carry out the thermodynamic analysis on the two thermal expansion processes.

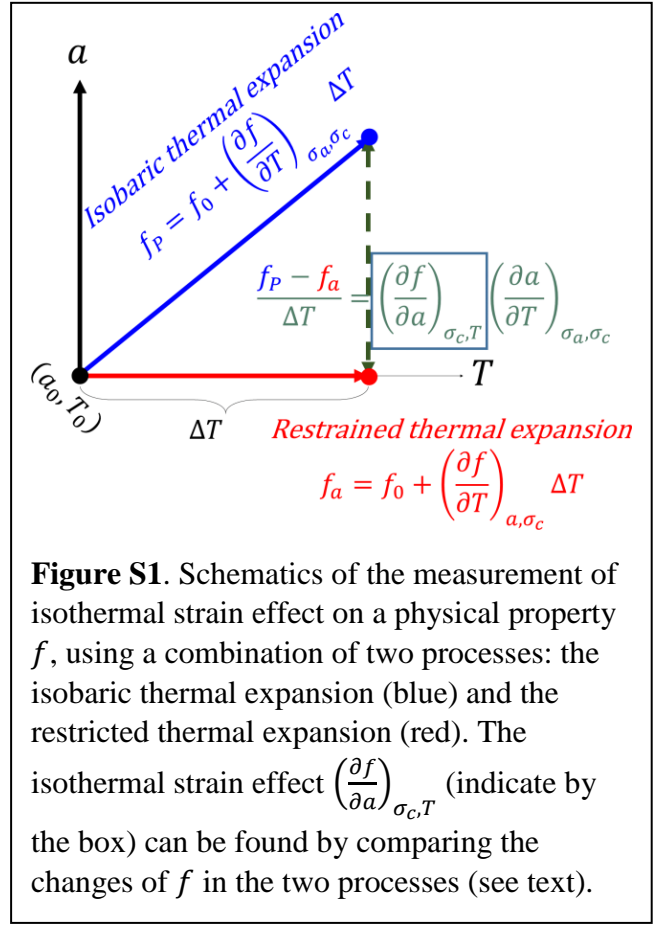

**Figure S1**. Schematics of the measurement of isothermal strain effect on a physical property $f$, using a combination of two processes: the isobaric thermal expansion (blue) and the restricted thermal expansion (red). The isothermal strain effect $\left(\frac{\partial f}{\partial a}\right)_{\sigma_c,T}$ (indicate by the box) can be found by comparing the changes of $f$ in the two processes (see text).

We start with the identification of the variable (state function) space. We choose three pairs of variables $(S, T), (\sigma_a, a), (\sigma_c, c)$, where $S$ is entropy, $T$ is temperature, $a$ and $c$ are lattice constants, $\sigma_a$ and $\sigma_c$ are the stress along the $a$ and $c$ directions. For simplicity, here we discuss biaxial strain ($\sigma_a = \sigma_b, a = b$). The treatment of more general strains can be carried out similarly.

The goal here is to find $\left(\frac{\partial f}{\partial a}\right)_{\sigma_c,T}$, i.e., the isothermal effect of strain along a certain direction ($a$), on another state function or physical property ($f$), while keeping the stress of the other direction ($\sigma_c$) constant.

It is in general difficult to measure $\left(\frac{\partial f}{\partial a}\right)_{\sigma_c,T}$ directly in experiments, since the experimental condition is tricky to realize, especially for brittle materials. On the other hand, experimentally, we can measure the change of $f$ in two processes of thermal expansion.

The first process is isobaric thermal expansion, with no change of stress at any direction, which measures the coefficient $\left(\frac{\partial f}{\partial T}\right)_{\sigma_a,\sigma_c}$.

The second process is the restrained thermal expansion, in which the lattice constant of the $a$ axis is kept constant; this measures $\left(\frac{\partial f}{\partial T}\right)_{a,\sigma_c}$. The initial conditions are $a_0, c_0, T_0$, and $f_0$.

After the isobaric thermal expansion, one has

$$f_P = f_0 + \left(\frac{\partial f}{\partial T}\right)_{\sigma_a,\sigma_c} \Delta T.$$

After the restrained thermal expansion, one has

$$f_a = f_0 + \left(\frac{\partial f}{\partial T}\right)_{a,\sigma_c} \Delta T.$$

It follows from that

$$f_P - f_a = \left(\frac{\partial f}{\partial T}\right)_{\sigma_a,\sigma_c} \Delta T - \left(\frac{\partial f}{\partial T}\right)_{a,\sigma_c} \Delta T,$$

or

$$\frac{f_P - f_a}{\Delta T} = \left(\frac{\partial f}{\partial T}\right)_{\sigma_a,\sigma_c} - \left(\frac{\partial f}{\partial T}\right)_{a,\sigma_c}.$$

Since

$$\left(\frac{\partial f}{\partial T}\right)_{\sigma_a,\sigma_c} = \left(\frac{\partial f}{\partial T}\right)_{a,\sigma_c} + \left(\frac{\partial f}{\partial a}\right)_{\sigma_c,T} \left(\frac{\partial a}{\partial T}\right)_{\sigma_a,\sigma_c},$$

one has

$$\frac{f_P - f_a}{\Delta T} = \left(\frac{\partial f}{\partial a}\right)_{\sigma_c,T} \left(\frac{\partial a}{\partial T}\right)_{\sigma_a,\sigma_c},$$

or

$$\left(\frac{\partial f}{\partial a}\right)_{\sigma_c,T} = \frac{f_P - f_a}{\Delta T} \frac{1}{\left(\frac{\partial a}{\partial T}\right)_{\sigma_a,\sigma_c}}.$$

We can measure $\left(\frac{\partial a}{\partial T}\right)_{\sigma_a,\sigma_c}$ in the isobaric thermal expansion experiments. Therefore, $\left(\frac{\partial f}{\partial a}\right)_{\sigma_c,T}$ can be measured using this two-process method.

## S2. Analysis of the time scale in thermal conduction using the model of RC analogy

### 2.1 RC analogy

A hot object may release its thermal energy and reduce its temperature by conducting heat to the environment; this is similar to the discharge of a capacitor.

The time scale of discharging a capacitor is

$$\tau_{RC} = RC,$$

where $C$ is the capacitance and $R$ is the resistance.

Similarly, one can estimate the time scale of the heat dissipation by thermal conduction using thermal conductance $G_h$ and heat capacity $C_h$: $\tau_{GC} = \frac{C_h}{G_h}$.

One can calculate the thermal conductance as $G_h = \frac{\sigma A}{d}$, where $\sigma$ is the thermal conductivity, $A$ is the area, and $d$ is the thickness of the material along the direction of the conduction. Note that here we assume that the thermal conduction is limited by the conduction in the material itself and the interfacial conductance is very high.

The heat capacity can be estimated as $C_h = c_M \rho A d$, where $c_M$ is the mass specific heat, and $\rho$ is mass density. Putting all together, one has:

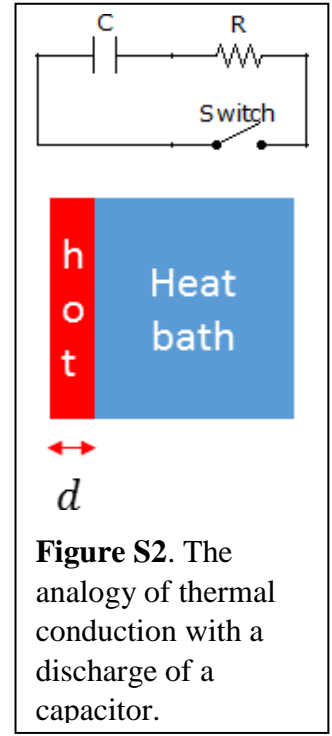

**Figure S2.** The analogy of thermal conduction with a discharge of a capacitor.

$$\tau_{RC} = \frac{C_h}{G_h} = \frac{c_M \rho A d}{\frac{\sigma A}{d}} = \frac{c_M \rho}{\sigma} d^2.$$

The decay time scales with the $d^2$. In other words, the thicker the material, the slower the heat dissipates.

### 2.2 Estimation of time scale of thermal conduction in h-LuFeO$_3$ and h-YFeO$_3$

Here we are concerned with the thermal conduction of h-LuFeO$_3$ (001) and h-YbFeO$_3$ (001) films to the substrates. Since the film thickness (~ 10 nm) is much smaller than the lateral dimension (~ 1 mm), we can treat this as a one dimensional (along the out-of-plane direction) problem.

For the h-LuFeO$_3$ and h-YbFeO$_3$ films measured in this work, the film thickness ($d$) is 25 and 50 nm respectively. Using the properties of h-LuFeO$_3$ and h-YbFeO$_3$ listed in the next section, we can estimate $\tau_{GC} = 0.6$ and 2.4 ns respectively, consistent with the time scale observed in the experiments (see Fig. S4).

## S3. Analysis of the time scale in thermal conduction using the diffusion model

### 3.1 Hot spot model

In general, a one-dimensional thermal conduction follows the diffusion equation [1]:

$$\frac{\partial T(x,t)}{\partial t} = D \frac{\partial^2 T(x,t)}{\partial x^2},$$

where $T$ is temperature, $D$ is the diffusion coefficient, $x$ is the spatial dimension and $t$ is time.

Depending on the initial and boundary conditions, there are many solutions. One solution is the following hot-spot equation:

$$T(x,t) = T_0 + \frac{A}{\sqrt{t+t_0}} \exp\left(-\frac{x^2}{4D(t+t_0)}\right),$$

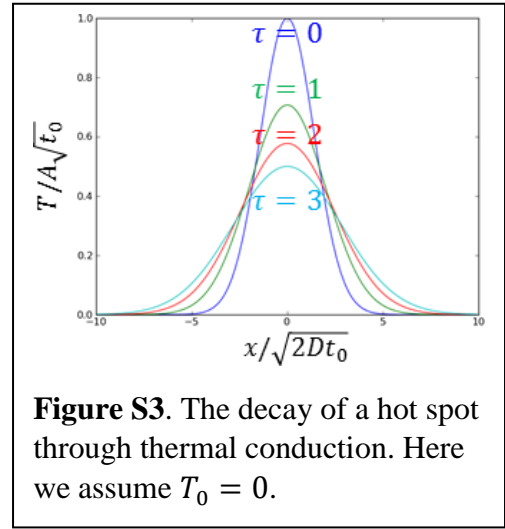

**Figure S3**. The decay of a hot spot through thermal conduction. Here we assume $T_0 = 0$.

where $A$ and $t_0$ determine the initial amplitude and width of a Gaussian distribution. This equation describes the decay of a hot spot, as a broadening of the Gaussian distribution over time. Defining the normalized time $\tau = t/t_0$, one can rewrite the equation into:

$$T(x,\tau) = T_0 + \frac{(A/\sqrt{t_0})}{\sqrt{1+\tau}} \exp\left(-\frac{x^2}{(4Dt_0)(1+\tau)}\right).$$

In terms of normalized time $\tau$, the relative changes of the width and amplitude are universal, as shown in Fig. S2. In other words, the decay of the Gaussian distribution scales with $t_0$. Since the initial width of the Gaussian is $w_0 = \sqrt{2Dt_0}$, then the decay time scales with $w_0^2$.

In other words, the decay time of a hot-spot defined by the relative change of the amplitude scales with the initial width of the distribution squared. The larger the spot, the slower it decays.

### 3.2 Fitting the experimental data on h-LuFeO3/Al2O3 and h-YbFeO3/YSZ films

More detailed analysis has to take into account the conduction in the film, in the substrate, and through the film/substrate interface. The equations for thermal conduction are the following:

$\rho c_M \frac{\partial T}{\partial t} = K \frac{\partial^2 T}{\partial z^2}$ (diffusion equation in the bulk of the film and subtrate), and $\rho c_M \frac{\partial T}{\partial t} = g \Delta T$ (diffusion equation at the interface), where $g$ is the thermal conductivity at the interface.

For the interface between the film and the air (film surface), we assume $g = 0$. On the other hand, the important parameter, the thermal conductivity of the film/substrate interface, is unknown. We therefore find the values by fitting the decay curve using $g$ as the fitting parameter.

**Table I** Thermal properties of the film and the substrate material used in the work. YSZ stands for yttrium stabilized ziconia. The densities of the materials are calcuated according to their unit cell size and composition. The mass specific heat is calcuated according to the Dulong Petite's law [2]. Thermal conducitivies of hexaongal ferrites (h-LuFeO$_3$ and h-YbFeO$_3$) has not been measured; the values are from the LuMnO$_3$ [3] which is isomorphic to the hexagonal ferrites.

| Symbol | Description | Unit | h-LuFeO$_3$ | h-YbFeO$_3$ | Al$_2$O$_3$ | YSZ |
|---|---|---|---|---|---|---|
| $\rho$ | Density | kg/m$^3$ | 9.1×10$^3$ [4] | 9.1×10$^3$ [5] | 3.9×10$^3$ [6] | 10.3×10$^3$ [7] |
| $\sigma$ | Thermal conductivity | W/(m·K) | 4 [3] | 4 [3] | 30 [8] | 2 [9] |
| $c_M$ | Mass specific heat | J/(kg·K) | 444 | 444 | 1215 | 352 |
| $D \equiv \dfrac{\sigma}{c_M \rho}$ | Diffusivity | m$^2$/s | 9.9×10$^{-7}$ | 9.9×10$^{-7}$ | 6.3×10$^{-6}$ | 5.5×10$^{-7}$ |

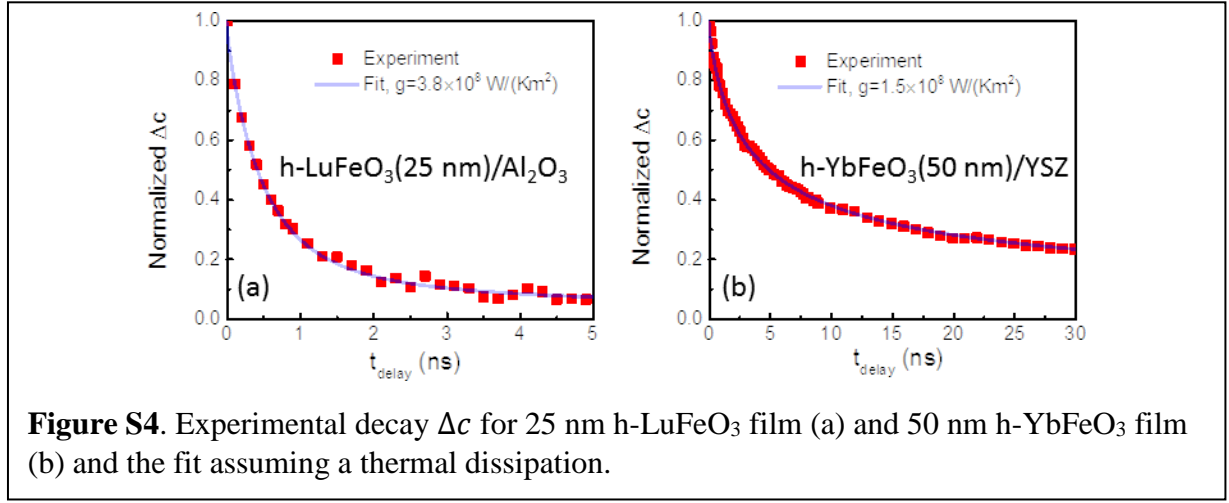

**Figure S4**. Experimental decay $\Delta c$ for 25 nm h-LuFeO$_3$ film (a) and 50 nm h-YbFeO$_3$ film (b) and the fit assuming a thermal dissipation.

We fit the experimental decay of the lattice constant using the interfacial thermal conductivity $g$ as the only fitting parameter. All the other parameters used are listed in **Table I**. As shown in Fig. S4, the single parameter $g$ can fit both measurements very well. In addition, the thermal conductivity appears to be higher in h-LuFeO$_3$/Al$_2$O$_3$ interface than that in the h-YbFeO$_3$/YSZ interface.

Notice that the areal thermal conductivity of the h-LuFeO$_3$ film and h-YbFeO$_3$ film can be estimated as $\sigma/d = 1.6\times10^8$ and $0.8\times10^8$ W/(K m$^2$) respectively, significantly smaller than the interfacial conductivity, justifying the estimation using the analogy of a discharging capacitor. Therefore, the slower decay of the lattice constant of the h-YbFeO$_3$ film can be attributed mostly to the larger film thickness, as discussed in the previous section.

## S4. Possibility of electronic origin for the photo-generated lattice expansion

When an above-band-gap photon is absorbed by the film, electrons are excited from the valence band to the conduction band, followed by a quick decay to the bottom of the conduction band by dissipating energy into the lattice (heating). After that, the absorbed energy exits the sample mainly in two ways: 1) by emitting photons generated in the electron-hole recombination; 2) by transmitting heat to the environment (substrate) via heat dissipation. So there are two possible scenarios for the observed lattice expansion. In the first scenario, the lattice expansion is a result of the electronic excitation, and the decay of the expansion corresponds to the decay of the electronically excited states. In the second scenario, the lattice expansion is a result of the temperature increase, while the decay of the expansion is due to the decrease of temperature via heat dissipation through the thermal conduction.

In the first scenario, the electronic excitation may affect the lattice constants in ferroelectric materials due to the piezoelectric effect. [10,11] In this regard, the photo-generated charge carriers in the conduction band may affect the lattice by screening the depolarization field generated by the ferroelectric polarization. This change of depolarization field changes the lattice constant along the polarization direction, due to the piezoelectric-like effect. Although the piezoelectric coefficient δ is unknown for hexagonal ferrites, we may get a rough estimate using the values of the isostructural hexagonal manganites, which is δ < 1 pm/V. [12] Assuming $P$ = 6.2 μC/cm² as the maximum polarization, [13] $\epsilon$ = 37 as the relative dielectric constant, [14,15] the maximum depolarization field can be found as $E = \frac{P}{\epsilon\epsilon_0} = 1.9 \times 10^8$ V/m. Assuming all the depolarization field is screened, which is very unlikely, [10] the relative change is about $\frac{\Delta c}{c} = E\delta < 1.9\times10^{-4}$. This value is much less than the observed magnitude which is up to $\frac{\Delta c}{c} = 4.4\times10^{-4}$, as shown in Fig. 2(b) in the main text. Therefore, the first scenario is unlikely to be majorly responsible for the observed lattice expansion.

A critical difference between the first and second scenario is the dependence on film thickness. In the first scenario, the time scale of the decay of the electronic state is not expected to be strongly dependent on the film thickness. In contrast, in the second scenario, the time constant is proportional to $d^2$. As shown in Fig. S4, a similar experiment on a 50 nm h-YbFeO₃ film shows a much longer time scale that can be explained using the model of thermal expansion, indicating that the second scenario is dominantly responsible for the lattice observed expansion.

## S5. Optical pump-probe experiment on the h-LuFeO$_3$ film

We have carried out the optical pump-probe experiment on a h-LuFeO$_3$/Al$_2$O$_3$ (80 nm) at room temperature to get insight on the time scale of the faster decay after the photon absorption. The sample was excited using a 400 nm laser pump, followed by an 800 nm laser pulse to measure the reflectance ($R$). As shown in Fig. S6, there appears to be fast decay within the first 100 ps after the laser excitation, which is followed by a much slower decay. The first decay could be related to electronic excitation, while the second could be related to the thermal dissipation.

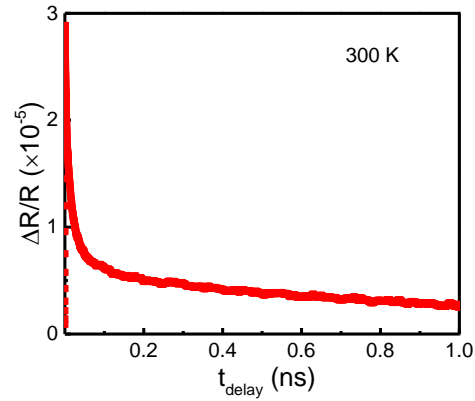

**Figure S6**. Optical pump-probe experiment showing the decay of reflectance as a function of time. The wavelength of the pump (probe) laser pulse is 400 (800) nm.

## S6. Comparison of the thermal expansion coefficients of h-LuFeO₃ with that of LuMnO₃

The thermal expansion in h-LuFeO₃ observed here is different from the behavior in bulk LuMnO₃ which shows a similar expansion in the basal plane as that in h-LuFeO₃ but a rather small contraction along the $c$ axis when the temperature is increased. [18] This difference may have to do with the electronic structure of the $Mn^{3+}$ that has empty $zz$ orbitals for the 3d electrons, which reduces the repulsion between the apex $O^{2-}$ and $Mn^{3+}$ center, causing less thermal expansion. In contrast, all the 3d orbitals in $Fe^{3+}$ is half-filled, making the repulsion between $Fe^{3+}$ and all the $O^{2-}$ atoms strong, generating similar thermal expansion coefficients in different crystalline directions.

## S7. Temperature change of the h-LuFeO$_3$ film after absorbing a photon pulse

The temperature change after absorbing a photon pulse can be found using the formula [10]

$$\Delta T = \frac{1}{c_M \rho d} \frac{E_p - E_g}{E_p} F_p A,$$

where $E_p$ is the photon energy, $E_g$ is the band gap, $F_p$ is the laser fluence, and $A$ is the absorbance. The absorbance can be calculated using $A = (1 - R)[1 - \exp(-d\alpha)]$, where $R$ is the reflectance, $\alpha$ is the absorption coefficient of h-LuFeO$_3$.

**Table II** Optical properties of h-LuFeO$_3$ and other parameters for calculating temperature change.

| Symbol | Description | Unit | Value |
|---|---|---|---|
| $E$ | Photon energy | eV | 3.18 |
| $E_g$ | Band gap of h-LuFeO$_3$ | eV | 2.0 [16,17] |
| $F_p$ | Laser fluence | mJ/mm$^2$ | 1.0 |
| $R$ | Reflectance at 3.1 eV (390 nm) | | 0.16 (measured) |
| $\alpha$ | Absorption coefficient at 3.1 eV | m$^{-1}$ | 7.5×10$^6$ [16,17] |
| $d$ | Film thickness | nm | 25 |

For the 25 nm h-LuFeO$_3$/Al$_2$O$_3$ film, the calculated (using absorption coefficient and reflectance) absorbance is 0.14, consistent with value 0.13 we meausured experimentally. Using absorbance $A$ = 0.13 and the parameters in Table II and Table I, one can estimate, a laser pulse of fluence 1 mJ/mm$^2$ increases the temperature by ≈ 460 K in the 25 nm h-LuFeO$_3$/Al$_2$O$_3$. The uncertainty is about 10%.

## S8. Debye-Waller factor in the temperature dependence of diffraction peaks

The temperature dependence of the diffraction peak intensity is affected by the thermal motion of the atoms via a factor $e^{-\frac{B}{2d^2_{(HKL)}}}$, where $B$ is called Debye-Waller factor and $d_{(HKL)}$ is the d-spacing of the $(H, K, L)$ diffraction plane (peak). In the case of $(1,0,L)$ peaks of h-LuFeO$_3$, considering both the Debye-Waller factor and the structure distortion, the peak intensities can be written as:

$$I_{(10L)} = I^0_{(10L)} \left(\frac{Q_{K3}}{Q^0_{K3}}\right)^2 e^{-\frac{B}{2d^2_{(10L)}}},$$

assuming that the oxygen contribution to the diffraction is ignorable.

Therefore, one has

$$\ln\left(\frac{I_{(10L)}}{I^0_{(10L)}}\right) = \ln\left(\frac{Q_{K3}}{Q^0_{K3}}\right)^2 - \frac{B}{2d^2_{(10L)}}$$

If we measure the temperature dependence of two peaks, say (102) and (106), and look at the ratio change between the two peaks, we can get

$$\ln\left(\frac{I_{(102)}}{I^0_{(102)}}\right) - \ln\left(\frac{I_{(106)}}{I^0_{(106)}}\right) = -\frac{B}{2d^2_{(102)}} + \frac{B}{2d^2_{(106)}} = \left(\frac{1}{2d^2_{(106)}} - \frac{1}{2d^2_{(102)}}\right) B.$$

In general, the Debye-Waller factor $B$ is a function of temperature, which can then be found by measuring the temperature dependence of intensities of at least two $(1,0,L)$ peaks.

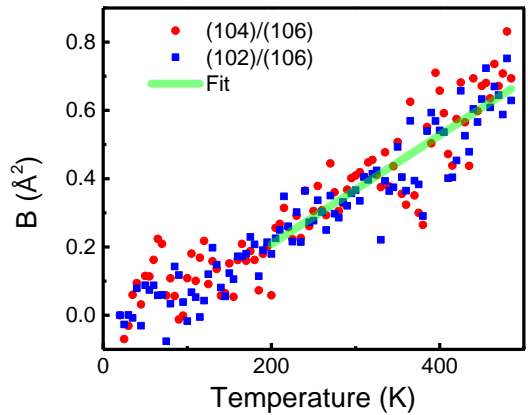

**Figure S5**. Debye-Waller factor as a function temperature in h-LuFeO3 calculated by comparing temperature dependence of (102), (104), and (106) peaks.

As shown in Fig. S5, we have measured temperature dependence of (102), (104), and (106) peaks. Using the relation discussed above, we calculated the Debye-Waller factor by comparing the temperature dependence of (102) and (106) peaks, and by comparing the temperature dependence of (104) and (106) peaks. The results found in the two comparisons match each other very well. Fitting the data in Fig. S5 ($T > 200$ K) using a linear function, one gets $B(T) = -0.11 + 0.0016T$.

In principle, one can use the Debye-Waller factors found here to renormalize the intensity data. In the isobaric thermal expansion, $B_P(T) = 8\pi^2 \langle u_x^2 \rangle$, where $u_x$ is the atomic motion along $\vec{q}$, considering the motion of atoms along all directions, $\langle u_x^2 \rangle = \frac{1}{3}\langle u^2 \rangle$. In the restrained thermal expansion, two direction of motion is restrained, so $B_a(T) = 8\pi^2 \langle u_x^2 \rangle = \frac{1}{3} 8\pi^2 \langle u^2 \rangle = \frac{1}{3} B_P(T)$. In other words, in the restrained thermal expansion, the Debye-Waller factor is smaller.